\documentclass[useAMS,usenatbib]{mn2e}
\usepackage{float,graphics,epsfig,array,amsmath,amssymb}
\usepackage{times}
\def\aj{AJ}%
\def\araa{ARA\&A}%
\def\apj{ApJ}%
\def\apjl{ApJ}%
\def\apjs{ApJS}%
\def\aap{A\&A}%
\def\aapr{A\&A~Rev.}%
\def\aaps{A\&AS}%
\def\mnras{MNRAS}%
\def\nar{New A Rev.}%
\def\pasj{PASJ}%
\def\nat{Nature}%
\def\memsai{Mem.~Soc.~Astron.~Italiana}%
\def\physrep{Phys.~Rep.}%
\newcommand{\ltsima}{$\; \buildrel < \over \sim \;$}
\newcommand{\simlt}{\lower.5ex\hbox{\ltsima}} 
\newcommand{\gtsima}{$\; \buildrel > \over \sim \;$}
\newcommand{\simgt}{\lower.5ex\hbox{\gtsima}} 

\newcommand{\xmm}{{XMM-\emph{Newton}}}

\newcommand{\asca}{{\emph{ASCA} }}
\newcommand{\lum}{erg~s$^{-1}$}
\newcommand{\flux}{{erg~cm$^{-2}$~s$^{-1}$ }}
\newcommand{\nh}{cm$^{-2}$}

\newcommand{\chandra}{{\emph{Chandra}}}

\newcommand{\sorg}{3C~445}
\newcommand{\suzaku}{{\emph{Suzaku}}}
\newcommand{\sax}{{\emph{BeppoSAX}}}

\newcommand{\swift}{{\emph{Swift}}}
\newcommand{\logxi}{erg cm s$^{-1}$}

\title[Suzaku deep observation of  3C~445]{Evidence for a circum-nuclear and ionised absorber in the  X-ray obscured Broad
Line Radio Galaxy 3C~445}
\author[Braito et al.]{V. Braito$^{1}$\thanks{e-mail address vb67@star.le.ac.uk}; J.N. Reeves$^{2}$; R.M.
Sambruna$^{3}$\thanks{mail address NASA HQ, 300 E Street SW, Washington, DC, 20546},J.
Gofford$^{2}$\\
 $^{1}$X-Ray Astronomy Observational Group, Department of Physics and Astronomy, Leicester University, Leicester LE1 7RH, UK\\
$^{2}$Astrophysics Group, School of Physical and Geographical Sciences, Keele University, Keele, Staffordshire ST5 5BG\\
$^{3}$Astrophysics Science Division, Mail Code 662, NASA Goddard Space Flight Center, Greenbelt, MD 20771, USA\\
}

\begin{document}

\date{}

\pagerange{\pageref{firstpage}--\pageref{lastpage}} 

\maketitle

\label{firstpage}

\begin{abstract}
Here we present the results of a   \suzaku\ observation of  the Broad Line Radio Galaxy \sorg.
We confirm the results obtained with the previous X-ray observations which unveiled  the
presence of several soft X-ray emission lines and an overall  X-ray emission which strongly
resembles a typical Seyfert 2 despite of the optical classification as an  unobscured AGN.    \\
The  broad band spectrum  allowed   us to  measure for the first time the  amount of reflection
($R\sim 0.9$) which together with the relatively strong neutral Fe K$\alpha$ emission line 
($EW\sim 100$ eV) strongly supports a scenario where a Compton-thick  mirror is present.     The
primary X-ray continuum is strongly obscured   by an absorber with a column density of
$N_{\mathrm H} =2-3 \times 10^{23}$\nh. Two possible scenarios are proposed for the absorber:  
a  neutral  partial covering  or a mildly ionised absorber with an ionisation parameter  $log
\xi \sim 1.0$ \logxi. A comparison with the past and more recent X-ray observations of \sorg\
performed with \xmm\ and \chandra\ is presented, which  provided tentative evidence  that the   
ionised and outflowing absorber varied.  We argue that the  absorber is probably associated with
an   equatorial   disk-wind located  within  the parsec scale molecular torus.
\end{abstract}


\section{Introduction}

 X-ray observations of AGN  are a powerful tool to understand the  physical conditions of the
matter in the proximity of the central SMBH and to understand the possible connection between
the   accretion and outflow mechanisms. In the last decade  X-ray observations have confirmed
the  widely accepted Unified Model \citep{Antonucci} of AGN, which accounts for the  difference
between  type 1 and type 2 AGNs through orientation effects.     However there are still several
open questions related to the geometry,  and physical state of the matter in the proximity of
the central supermassive black hole (SMBH).  In particular a   key question  still to be
answered  is the origin of  powerful relativistic jets and outflowing winds. Understanding the
nature of AGN with powerful jets or disk winds is not only important in order to understand the
accretion itself, but more importantly  the mechanism with which the AGN can expel the gas
supply of   the host galaxy quenching the growth of the SMBH and of the galaxy itself. Indeed, 
jets and outflows  can transport a significant fraction of the mass-energy  into the AGN
environment and they could represent a key to understand the AGN feedback mechanism
\citep{Fabian2010,Elvis06,Cattaneo09, King03}. \\

In the last decade X-ray observations have   successfully revealed the presence of both warm and
cold gas in the central region of AGN, and  have shown that   $\sim50$\% of the X-ray spectra of
radio quiet AGN  (RQ-AGN) present  absorption and emission features due to the presence of
photoionised gas, which  is outflowing with   typical   velocities   of $\sim 100-1000$ km
s$^{-1}$ \citep{Crenshaw2003}.  On the other end, previous observations of  Radio-Loud AGN 
(RL-AGN) suggested that their X-ray emission is similar to the case of RQ AGN but with some  
differences. In particular the X-ray emission appears to be harder (or flatter) and with weaker
features   due to reprocessing (reflection/absorption) from warm or cold gas with respect to the
RQ population \citep{S02,Grandi2006, Balla}.  Several possible scenarios were proposed  to
account for these differences, among them  a smaller subtending angle of the reprocessing
medium  as in an advection dominated accretion-flow (ADAF) models \citep{ADAF} or a higher
ionisation state of the accretion disk \citep{Balla2002}. \\ 

Recent sensitive and broadband observations performed with  \chandra, \xmm\ and  \suzaku\ are
subverting this view, indeed observations of samples of RQ and RL AGN have shown a large
variety  in the X-ray properties, with a wider range of X-ray continuum shapes  where the RL
sources  simply populate one end of the distribution. Emission and absorption lines have been
now detected  also in the soft X-ray spectra of Radio Galaxies
(\citealt{Torresi2010,445chandra,Reeves2009,Torresi2009,S07}; hereafter   S07), indicating  the
presence of photoionised gas  in the central regions of RL-AGN analogous to RQ-AGN. Furthermore
a recent analysis of the  \suzaku\ observations of broad line radio galaxies (BLRGs) has also
unveiled the presence of fast outflowing gas with velocities $v\sim0.1-0.3 c$ and carrying
substantial masses and kinetic powers similar to the radio jets \citep{Tombesi2010}.  \\
 
We are currently carrying out a program of \suzaku\ observations of a sample of all the nearby
($z<0.1$) broad line radio galaxies (BLRGs), for which we also have \xmm\ and or \chandra\ data
(3C~390.3; \citealt{Sambruna2009}, 3C~111; Ballo et al. in prep; 3C~382; Sambruna et al. in
prep  and \sorg; \citealt{445chandra}). One  of the goals of this project is to investigate the
structure of the accretion flow and the presence of warm and cold gas in the central regions of
BLRGs to better understand the jet formation. Here we focus on one of these sources \sorg, while
the X-ray  properties of the sample will be presented in a forthcoming paper (Sambruna et al. in
prep).    

\subsection{The Broad Line Radio Galaxy \sorg.}
\sorg\ is a nearby (z=0.057) BLRG with a FRII morphology \citep{Kronberg}.  The
optical spectrum  of \sorg\  shows the presence, in total flux, of broad emission
lines (H$\alpha$ FWHM $\sim$6400 km s$^{-1}$; \citealp{Eracleous94})  typical of a
type 1 or unobscured AGN leading to  the classification as a BLRG. The optical
continuum is also highly reddened; from the  large Balmer decrement
(H$\alpha$/H$\beta \sim$ 8)  \citealp{Crenshaw88} derived   E(B-V)=1 mag,
consistent with the large Pa$\alpha$/H$\beta$ ratio (5.6; \citealp{Rudy}). 
Assuming a standard dust-to-gas ratio this corresponds to a column density  of N$_H
\sim 5 \times 10^{21}$ \nh, which is    one order of magnitude larger than the
Galactic column density in the direction to \sorg\ (N$^{Gal}_H=5.33 \times 10^{20}$
\nh, \citealp{Murphy96}).  From radio observations \citep{Eracleous98}  an 
inclination  for the jet of   $i>60^{\circ }$ is  inferred, suggesting that  the
contribution of the jet is negligible  (Doppler factor of $\delta\sim 0.2$) and 
also that \sorg\ is seen almost edge-on. A large viewing inclination angle for
\sorg\ ($i \leq 60^{\circ }$)  was  also inferred by  \citet{Grandi07}  with the
analysis of  the flux ratio of the jet  components  and the  VLA maps  presented by
\citet{Leahy}.\\

\sorg\ is a bright X-ray source (F$_{2-10~keV} \sim 7 \times 10^{-12}$ \flux),  and
was previously observed with all the major X-ray observatories. The analysis of 
an  archival 15 ks \xmm\ observation of \sorg\  showed a remarkable spectrum  (S07;
\citealt{Grandi07}). The intriguing result was  that, despite its optical
classification as a type 1 AGN, its  X-ray emission was typical of an obscured AGN
in several aspects.  The 2--10 keV continuum could be described by a heavily
absorbed (N$_H \sim 10^{23}$ \nh) power-law with photon index $\Gamma\sim 1.4$; a 
narrow and unresolved Fe K$\alpha$ emission line was also detected with  EW $\sim$
120 eV. Due to the limited EPIC bandpass (0.4--10 keV)  it was  impossible to
distinguish between a scenario with a strong reflection component (with $R\sim 2$)
or a multi-layer neutral partial covering absorber. However, in both the scenarios
the    measured $N_\mathrm{H}$  exceeded  the column density  predicted from the
optical reddening.      Furthermore,  the soft X-ray spectrum was also  reminiscent
of a Compton-thin Seyfert 2, with several   emission lines in 0.6--3~keV, due to
ionised elements from O to Si. \\

At high energies \sorg\ was detected with the  \sax-PDS instrument
\citep{Grandi2006}. However, due to the large field of view of the  high energy
detector onboard \sax\  and the possible contamination from a nearby cluster
(A2440) this observation could provide only a weakly constrained  measurement of
the     reflected continuum.  \sorg\ is  also detected    with the BAT detector on
board Swift  and is part of the 58 months
catalogue\footnote{heasarc.gsfc.nasa.gov/docs/swift/results/bs58mon/} 
(\citealt{Baumgartner}). \\

In the soft X-ray band the  high  resolution spectra, accumulated with the 
\xmm-RGS,  provided the first  tentative detection of the {O\,\textsc{vii}}  and
{O\,\textsc{viii}}  emission lines  \citep{Grandi07}. This detection   suggested
that the soft X-ray emission  is produced in a warm gas photoionised by the AGN as 
observed in Compton-thin Seyfert 2  galaxies \citep{Bianchi06,Guainazzi07}.
Recently we obtained a deep \chandra\  observation of \sorg, which  provided  the
evidence for  both emission and absorption  from  photoionised gas in this obscured
AGN  and provided the first detailed quantitative measurement. In particular, in
the \chandra\ spectrum several  soft X-ray emission lines   (from    the He and
H-like transitions of O, Ne, Mg and Si) were detected and resolved. From the  ratio
of forbidden to intercombination emission lines in the He-like triplets and the
velocity broadening  ($v_{\mathrm{FWHM}}\sim 2600$ km s$^{-1}$) it was inferred
that the photoionised emitter has  properties very similar to   the Broad Line
Region. The \chandra\ spectrum confirmed that \sorg\ is  highly obscured but it was
suggestive that  X-ray absorber could be  associated with a   disk wind with an
observed outflow velocity of  10000 km s$^{-1}$, which is launched  from a
sub-parsec scale location.\\

Here we present  the analysis and results from  our 140 ks \suzaku\ observation of
\sorg, while the results on the  high resolution soft X-ray spectra obtained with
our deep (200 ksec) \chandra\ LETG observation are discussed in a companion paper
\citep{445chandra}.  The \swift-BAT spectrum ($S/N= 16.2$; $F(14-195)=4.2\pm0.5
\times 10^{-11}$) is  used in this paper and fitted jointly with the \suzaku\
data.  We took advantage of the   \chandra\  results,     while modelling the
\suzaku\ data (\S 3.2 and \S3.4), and   we present a  comparison  between the two
datasets, which  provided tentative evidence  that the    X-ray absorbers varied
(\S 3.4 \& \S4). \\

The paper is structured as follows. The   observation  and data reduction are
summarised in \S~2. In \S~3 we present the  modelling of the broad-band spectrum,
aimed to assess the  nature of the X-ray absorber,  the  amount of reflection and 
the  iron K emission line  properties.  Discussion and conclusions follow in \S\S~4
and 5. Throughout this paper, a concordance cosmology with H$_0=71$ km s$^{-1}$
Mpc$^{-1}$, $\Omega_{\Lambda}$=0.73, and $\Omega_m$=0.27 \citep{Spergel2003} is
adopted. 

\section{Observations and data reduction}
 On  25 May 2007   \suzaku\    \citep{Mitsuda07} observed \sorg\  for a total
exposure time of about 140~ksec (over a total duration  of $\sim 270$ ksec); a
summary of observations is  shown in Table\,\ref{table:log_observ}.   \suzaku\  
carries on board  four sets of X-ray mirrors, with an X-ray CCD  (XIS; three front
illuminated, FI\footnote{At the time of this observation  only XIS0 and XIS3   were
still operating},  and one back illuminated, BI) at their focal plane, and a non
imaging hard X-ray detector (HXD).  All together the XIS and the HXD-PIN cover the
0.5--10 keV and 12--70 keV bands respectively. Data from the XIS
(\citealp{Koyama07}) and HXD-PIN \citep{Takahashi} were  processed using v2.1.6.14
of the \suzaku\ pipeline  applying the standard screening\footnote{The screening
filters all  events  within the South Atlantic Anomaly (SAA)  as well as  with an
Earth elevation angle (ELV) $ < 5\ensuremath {{}^{\circ }}$ and  Earth day-time
elevation angles (DYE\_ELV) less than $ 20\ensuremath {{}^{\circ }}$. Furthermore
also data within  256s of the SAA were excluded from the XIS and within 500s of the
SAA for the HXD. Cut-off rigidity (COR) criteria of $ > 8 \,\mathrm{GV}$ for the
HXD data and $ > 6 \,\mathrm{GV}$ for the XIS were used.}. 

\subsection{The Suzaku-XIS analysis}
The XIS data were selected in $3 \times 3$ and $5\times 5$ editmodes using only
good events with grades 0,2,3,4,6 and filtering the  hot and flickering pixels with
the script \textit{sisclean}. The net exposure times are   $108$ ksec  for each of
the XIS. The XIS  source spectra  were extracted from a circular region of 2.4$'$
radius  centered on the source,  while background spectra  were extracted from two
circular regions of 2.4$'$ radius  offset from the source and the calibration
sources.  The XIS response (rmfs) and ancillary response (arfs) files were
produced,   using the latest calibration files available, with the \textit{ftools}
tasks \textit{xisrmfgen} and \textit{xissimarfgen} respectively.  The spectra from
the  two FI  CDDs (XIS 0 and XIS 3) were combined to create  a single source
spectrum (hereafter XIS--FI), while the  BI (the XIS1) spectrum  was kept separate
and fitted simultaneously.    The net 0.5--10 keV  count rates  are: $(0.154\pm
0.001)$ cts/s, $(0.158\pm 0.001)$ cts/s, $(0.142\pm 0.001)$ cts/s for the  XIS0,
XIS3 and XIS1  respectively with a net exposure of 108 ks for each XIS.    Data
were included from 0.4--10 keV for the XIS--FI and  0.4--8 keV for the XIS1 chip;
the difference on the upper boundary for the XIS1 spectra is because this CCD is
optimised for the  soft X-ray band. The background rates  in these energy ranges
correspond to only 4.5\% and 8.1\% of the net source counts   for the XIS-FI and
XIS1 respectively.  The net XIS source spectra were then  binned   to a  minimum 
of 100 counts per bin   and   $\chi^2$ statistics have been used. 

\subsection{The \suzaku\ HXD-PIN analysis}
For the HXD-PIN data  reduction and analysis we followed the latest \suzaku\ data
reduction guide (the ABC guide Version
2\footnote{http://heasarc.gsfc.nasa.gov/docs/suzaku/analysis/abc/}),   and used the
rev2 data, which include all 4 cluster units.   The HXD-PIN instrument team
provides the background (known as  the ``tuned'' background) event file, which
accounts for  the instrumental  Non X-ray Background (NXB; \citealp{kokubun}).  The
systematic uncertainty of   this ``tuned'' background model is believed to be 
±1.3\% (at the 1$\sigma$ level for a net 20 ks exposure).\\   We extracted the
source and background spectra   using the same common good time interval, and 
corrected the source spectrum  for the detector dead time. The net exposure time 
after  the screening  was 109.5 ks. We  then simulated a spectrum for the cosmic
X-ray background counts \citep{Boldt,Gruber} and added  it to the  instrumental
one.\\  

\sorg\ is detected  up to 70 keV at a level of 12.9\% above the background 
corresponding to a  signal-to noise ratio $S/N\sim 30$. The net count rate in the
15--70 keV band is $0.052\pm 0.002$cts/s ($\sim5700 $ net counts).  For the
spectral  analysis the source  spectrum of \sorg\ was rebinned   in order to have a
signal-to-noise ratio of  five in each energy bin. The HXD-PIN spectrum can be
fitted with a single power-law ($\Gamma=2.0\pm0.3$); this provides a first estimate
of the 15-70 keV flux of about $\sim 3.1\times 10^{-11}$\flux.  The extrapolated
flux in the \swift\  band (14-195 keV) is  $\sim 5\times 10^{-11}$\flux and it is
comparable to the flux reported in the BAT 58 months catalog
(\citealt{Baumgartner}).

\subsection{The \swift-BAT observation} 
The BAT spectrum was obtained  from the 58-month  survey archive, which provides
both the spectrum and  the long-term online BAT light curve; the data reduction
procedure of the eight-channel spectrum  is described in (\citealt{Tueller10} and
\citealt{Baumgartner}).   For the analysis we used the  latest calibration
response  file (the diagonal matrix: diagonal.rsp) provided with the spectrum and
we  also inspected the light curve  that shows no strong variability. The net  
count rate in the 14--195 keV band   is  $(4.2 \pm 0.3)\times 10^{-4}$ cts
s$^{-1}$.

\subsection{The \chandra\ observation}
Recently \sorg\ was observed with the \chandra\ ACIS-S  for 200 ksec. The
observation  was performed on September 2009 with the Low-Energy Transmission
Grating (or LETG; \citealt{letg_ref}) in the focal plane.  In this paper we 
concentrate on  the \suzaku\ results, while the \chandra\ data reduction, analysis
and results are described in a companion paper \citep{445chandra}.  Thanks to the
high   sensitivity and spectral resolution the LETG  data allowed us to resolve the
soft X-ray emission lines, determining the gas density and location. Since \sorg\
is not highly variable  either in flux and spectral shape, we were able to  take
the advantage of the \chandra\ results, while modelling the \suzaku\ data and
vice-versa.

\begin{table}
\caption{Summary of  the observations used: Observatory, Epoch, Instrument, Total and Net exposure time.
For \suzaku\ the total exposure time  is not the elapsed time of the observation and it already includes
the standard  screening for the passage above the SSA anomaly. The  net exposure times    are
after the screening of the cleaned event files.  \label{table:log_observ}
} 
\begin{tabular}{ccccc}
 \hline
 Mission &  DATE & Instrument  & T$_{\rm(tot)}$ (ks)&T$_{\rm(net)}$ (ks)\\
 \hline

\suzaku\ & 2007-5-25   &XIS-FI & 139.8& 108\\
 \suzaku &  2007-5-25 & XIS 1& 139.8& 108\\
 \suzaku & 2007-5-25 & HXD-PIN & 139.8 & 109.5\\
\chandra& 2009-09-25 &ACIS-S LETG & --&43.950\\
\chandra& 2009-09-29 & ACIS-S LETG & --&73.720\\
  \chandra& 2009-10-02 &ACIS-S LETG & --&83.440\\

 \hline
\end{tabular}
\end{table}

\section{Spectral Analysis}
All the models have been fit to the data using  standard software packages (XSPEC
ver. 11.3).   In the following, unless otherwise stated, fit parameters are quoted
in the rest frame of the source  and errors are at the 90\% confidence level   for
one interesting parameter ($\Delta\chi^2=2.71$).

\subsection{The broad band continuum} 
Previous X-ray studies of \sorg\ revealed that its X-ray spectrum is complex and
cannot be modelled with a single power-law component. This is confirmed by our
\suzaku\ observation, where the XIS  curved  continuum is highly indicative of the
presence of  strong absorption. A fit  to the 0.4--10 keV XIS data  with  a single 
redshifted power-law  model  modified by Galactic (N$_{\rm H}= 5.33\times
10^{20}$cm$^{-2}$; \citealp{Dickey1990})  and intrinsic (in the rest-frame of
\sorg) absorption, yields a poor fit ($\chi^2/dof=3865.2/388$), with a hard photon
index ($\Gamma\sim -0.27$) and leaves strong residuals  at low and high energies.\\

We then added to this model a soft power-law component absorbed only by the
Galactic column, which represents the  primary X-ray emission scattered into our
line of sight. The photon indices of these two power-law components were
constrained to be the same.   This model is still a poor description   the X-ray
emission from  \sorg\ ($\chi^2/dof=885.3/387$) and as already noted with the \xmm\
observation (S07) it fails to reproduce the overall curvature, which suggests that
a more complex absorber is required. Furthermore, it leaves strong residuals  both 
at the Fe K$\alpha$ line energy range and in the soft X-ray band  and  the photon
index is  found to be   hard   ($\Gamma \sim 1.3$)  with respect to the typical
values of radio loud AGN (\citealp{S99,Reeves00}).\\

We   then tested a model  for the continuum similar to the best fit found for the 
\xmm\  data, without including any reflection,  and we ignored the 5--7.5 keV band,
where the Fe K$\alpha$ emission complex is expected. The absorber is now modelled
with a dual absorber; one fully covering the primary X-ray emission and one only
partially covering it.  A scattered component, modelled with a second power-law
with the same photon index of the primary one is still included.  This is now a
better representation of the X-ray continuum and  the photon  index is now steeper
($\Gamma=1.70\pm 0.11$). The  column densities of the absorbers are found to be
N$_\mathrm{H1} = (1.1\pm  0.2)\times10^{23}$ \nh   and N$_\mathrm{H2}   = (3.3
\pm0.6)\times 10^{23}$ \nh, for the fully and partial covering absorber
respectively; the covering fraction  of the latter absorber is  $f_\mathrm{cov} =
0.79^{+0.06}_{-0.08}$, while the scattering fraction is found to be
$f_\mathrm{scatt} \sim 0.03$.\\

\begin{figure}
\begin{center}
\resizebox{0.46\textwidth}{!}{
\rotatebox{-90}{
\includegraphics{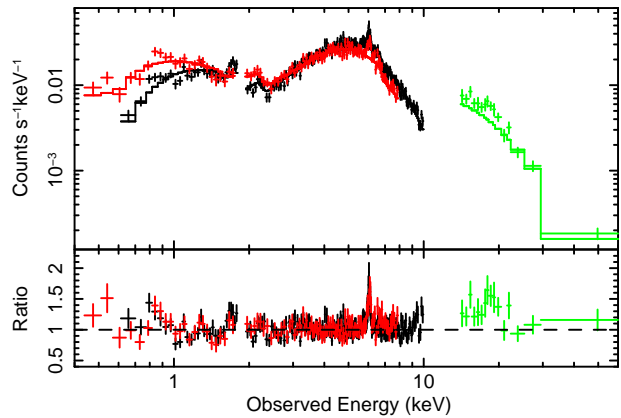}
}}
\caption{Suzaku 0.4--60 keV   spectra (in the electronic version: XIS-FI, black;
XIS1, red; HXD-PIN   green) of \sorg; data have been rebinned for plotting
purposes.  The upper panel shows the data and the dual absorber model ($\Gamma\sim
1.7$; N$_\mathrm{H1} \sim 10^{23}$ \nh;    N$_\mathrm{H2} \sim 3.3 \times10^{23}$
\nh), fitted over the 0.4--5 keV and 7.5--10 keV band. The lower panel shows the
data/model ratio  to this model. Clear residuals are visible at the   iron K-shell
energy band and in the HXD-PIN energy range.  The   excess of counts   in the
HXD-PIN  spectrum compared to the XIS,  is probably due to the Compton reflection
hump.
\label{fig:residualshxd.ps}
}
\end{center}
\end{figure}

\begin{figure}
\begin{center}
 \resizebox{0.5\textwidth}{!}{
\rotatebox{-90}{
\includegraphics{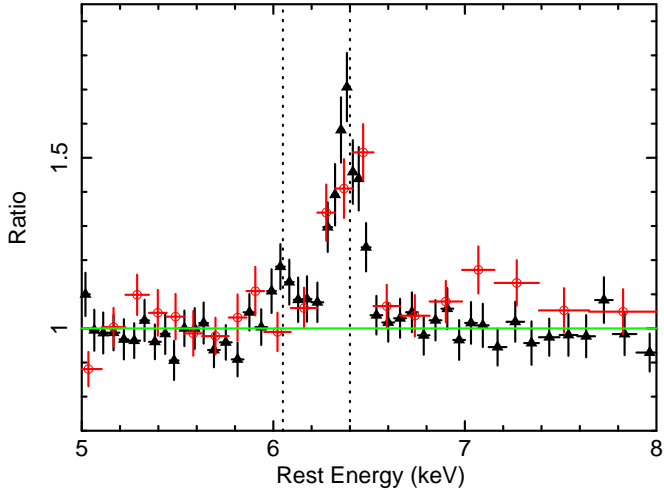}
}}
\caption{Data/model ratio between  the XIS data (XIS-FI, black  triangles in the
electronic version; XIS-BI red open circles in the electronic version) and the 
dual absorber model   showing the iron line profile. The data clearly show
(indicated by the two vertical lines) a narrow core at 6.4 keV (rest frame),  and a
weak   narrow emission feature at$\sim 6.1$ keV (rest frame). 
\label{fig:fe_line.ps}
}
\end{center}
\end{figure}

This continuum model is  still too simple with respect to the broad band X-ray
emission, indeed  its extrapolation  under-predicts the counts collected above 10
keV  (see Fig.~\ref{fig:residualshxd.ps}). Furthermore, as seen with \xmm\ this
model leaves  strong line-like residuals at  the energy of  the Fe K$\alpha$
emission line.  These residuals suggest  the presence of a   strong narrow core at
the expected energy of the Fe K$\alpha$  line (6.4 keV)   and a possible weak
component red-ward  the narrow core at $E\sim 6 $ keV (observed frame), which could
be identified with the  Compton shoulder (see Fig.~\ref{fig:fe_line.ps}).  Both
these features and   the hard excess  seen in the spectrum above 10 keV   suggest 
the presence of a strong  reflection component. The presence of this latter
component  was already  suggested   with the previous \sax\ observation
\citep{Dadina2007,Grandi2006}, however  taking into account  the  possible
contamination from a nearby cluster of galaxies (A2440, z=0.094) it was not
possible to derive strong constraints on it.   \\

In order to obtain a  better representation of the X-ray continuum,   we then 
fitted simultaneously  the    \suzaku\ XIS (0.4--10 keV)  and HXD-PIN data (15.--
65. keV) and the \swift-BAT   spectra. We set  the cross-normalization factor 
between the HXD and the XIS-FI spectra  to 1.16, as recommended for XIS nominal
observations processed after 2008 July (Manabu et al. 2007; Maeda et al.
2008\footnote{http://www.astro.isas.jaxa.jp/suzaku/doc/suzakumemo/suzakumemo-2007-11.pdf;\\
http://www.astro.isas.jaxa.jp/suzaku/doc/suzakumemo/suzakumemo-2008-06.pdf}), and
allowed the   cross-normalization of the \swift-BAT data   to vary, since the two
observations are not simultaneous.   \\ We included in the model the Fe K$\alpha$
line and a Compton reflection component.  At this stage this component was   
modelled   with  the  \,\textsc{pexrav} model in Xspec (\citealp{pexrav}), with 
the  abundances  set to solar values and the inclination angle $i $ to 60$^\circ$.
We note that the cluster A2440, which was thought to contaminate the X-ray emission
detected with the \sax-PDS instrument lies at the edge of the FOV of the HXD-PIN
and thus the contamination from it should be minimal. The cross-normalization with
the Swift data    is consistent with one as indicated by the similar HXD and BAT
fluxes ($C\sim 0.8$), furthermore the slope of the \swift-BAT spectrum is
consistent with the HXD-PIN data, with no evidence of a high-energy cutoff. The
similarity in flux and shape confirm that the \suzaku\ HXD-PIN spectrum is
dominated by  the emission of \sorg\  with no or minimal contamination from the
nearby cluster. It is worth noting that since the \swift-BAT data allow us  to
extend the analysis only up to 150 keV, we cannot discriminate if the lack of any
roll over is real or simply due to the still limited bandpass and the complex
curvature of the spectrum. Indeed upon leaving the  high energy cutoff free to vary
we can set only a lower limit ($E>60$ keV). \\

The  amount of reflection, defined   by the subtending solid angle of the reflector
$R=\Omega/2\pi$ is found to be $R\sim1.1\pm0.4$, while the parameters of the
absorbers are  consistent with the values obtained with the previous model
(N$_\mathrm{H1} = 1.1^{+0.1}_{-0.1}\times10^{23}$ \nh\  and N$_\mathrm{H2} =
3.2^{+0.4}_{-0.4}\times10^{23}$ \nh,  $f_\mathrm{cov} = 0.74^{+0.02}_{-0.02}$  and
$\chi^2/dof= 496.7/408$). The photon index is now $\Gamma=1.78^{+0.08}_{-0.07}$.  The Fe $ \mathrm{K}\alpha$ emission line is centered at $E=6.384 \pm 0.012
\,\mathrm{keV}$,    it  has an equivalent width of $EW= 105\pm 15\,\mathrm{eV}$
with respect to the observed continuum and it is   has a measured width of $ \sigma
= 37 \pm 34 \,\mathrm{eV}$.  We note that the  inclusion of the   reflection
component  is not only statistically required ($\Delta \chi^2=23$ for 1 dof or
$\chi^2/dof= 520.1/409$), but  also its strength is consistent  with  the
observed EW of the Fe $ \mathrm{K}\alpha$  emission line. This  model gives a 2--10
keV observed  flux  of $\sim 7\times 10^{-12}$ \flux and a intrinsic (corrected for
absorption) luminosity of $\sim 1.2\times 10^{44}$ \lum, which is similar to the
value measured with \xmm\ and \asca.  \\

This model now provides  a better phenomenological  description \sorg's X-ray
continuum, however statistically the fit is still poor ($\chi^2/dof= 496.7/408$),
with   some  residuals in the 2--10 keV band, suggesting that this model is still
too approximate for the broad band emission of \sorg.  Several line like residuals
are present at $E<2$ keV,  in agreement with   the   previous detection  from 
\asca and \xmm\ (\citealp{S99,S07,Grandi07}). Finally we note that without
the inclusion of the \swift-BAT, the overall  statistics ($\chi^2/dof= 491.7/401$)
and  parameters, derived with this simple and phenomenological model, are similar
($\Gamma=1.79^{+0.09}_{-0.08}$ and  $R\sim1.2\pm0.5$). The reflection component is
also statistically required  with the \suzaku\ data alone ($\Delta\chi^2 = 21 $ for
1 dof or $ \chi^2/dof= 513/402$). A more detailed description of the modelling  of
the  Fe  K$\alpha$ emission line and of the overall hard X-ray spectral curvature
is provided in the sections 3.3 and 3.4; nevertheless we note that  this modelling
does not strongly affect the results of the soft  X-ray  emission, which are
presented in the following section.  \\

 \begin{table*}
\caption{Summary of the soft-X-ray emission lines. The energies of the lines are  
quoted in the rest frame. Fluxes and possible identifications are reported in
column  2 and 3. For the emission feature detected at $\sim 0.88$ keV the
alternative identifications are reported in brackets. The EW are reported in column
4 and they are  calculated against the total  observed continuum at their
respective energies. In column 5 the improvement of fit is shown; the value for the
model with no lines is $\chi^2/dof=746.4/676$. In column 6 we report the
theoretical value for the transitions. The $\Gamma$ of the soft power-law is tied
to the  hard power-law component.  We note that  some of the soft X-ray emission
lines are not detected with high statistical significance (e.g  for {Ne
\,\textsc{x}} Ly$\alpha$ and  {Mg\,\textsc{xi}} He$\alpha$    we have $\Delta
\chi^2=7.7$ and
$\Delta\chi^2=5.2$, respectively).
\label{table:soft_lines} }
\begin{tabular}{cccccc} \hline

 Energy      & Flux  &ID        & EW & $\Delta \chi^2$ & E$_{\rm {Lab}}$ \\
 (keV)       & ($10^{-6}$ph cm$^{-2}$ s$^{-1}$) &(eV) &(eV) &&(keV) \\
 
    (1)   &  (2)   &  (3)    &  (4)   & (5)  & (6)  \\

\hline
     &     &      &     &    &  \\
0.54$^{+0.03}_{-0.02}$& 15.92$^{+0.72}_{-0.70}$&{O\,\textsc{vii}} He$\alpha$  &47.6$^{+22.3}_{-20.6}$& 10.85& 0.561(f); 0.569 (i); 0.574(r) \\   
   &   &   &     & &\\
0.88$^{+0.02}_{-0.02}$& 7.37$^{+1.98}_{-2.22}$& {O \,\textsc{viii}} RRC &51.0$^{+13.8}_{-14.5}$& 30.0& $>0.873$\\   
     &   &     ({Fe\,\textsc{xviii}-\,\textsc{xix} }) &  &&0.853-0.926 \\
 &   &   &     & &\\

0.99$^{+0.01}_{-0.03}$& 2.85$^{+1.45}_{-1.62}$& {Ne \,\textsc{x}} Ly$\alpha$&24.0$^{+12.2}_{-13.6}$&7.7&1.022\\
  &   &   &     & &\\
1.36$^{+0.03}_{-0.03}$&1.40$^{+0.78}_{-1.14}$& {Mg\,\textsc{xi}} He$\alpha$ & 19.5$^{+11.0}_{-15.8}$ &5.2&1.331(f); 1.343(i); 1.352 (r)\\
   &   &   &     & &\\
1.80$^{+0.03}_{-0.03}$& 2.89$^{+1.31}_{-0.46}$& {Si\,\textsc{xiii}} He$\alpha$ &59.9$^{+27.1}_{-9.6}$&20.7 & 1.839(f); 1.853(i); 1.867 (r)\\
    &   &   &     & &\\
 2.33$^{+0.05}_{-0.04}$ & 1.65$^{+0.76}_{-0.75}$ &  {S\,\textsc{i}} K$\alpha$&45.7$^{+21.1}_{-20.7}$&9.2& 2.307\\
     &     &      &     &    &  \\
 \hline
\end{tabular}
\end{table*}

 \subsection{The soft X-ray spectrum}
\begin{figure}
\begin{center}
\resizebox{0.48\textwidth}{!}{
\rotatebox{-90}{
\includegraphics{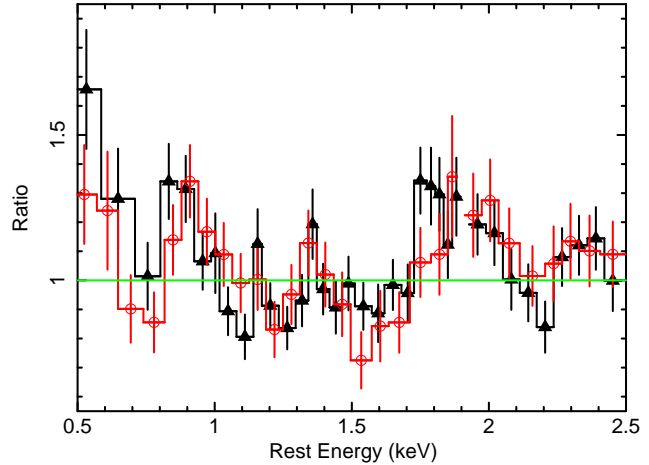}
}}
\caption{Zoom on  the 0.5--2.5 keV range of the data/model ratio to a power-law
component ($\Gamma\sim 1.7$; XIS-FI,   black  triangles in the electronic version;
XIS-BI open red circles in the electronic version).  Several emission lines are
clearly present in the residuals.  
\label{fig:soft.ps}
}
\end{center}
\end{figure}

In order to  model the soft X-ray emission, we   allow the photon index of the soft
component   to vary.  The  photon index of the scattered  component is now found to
be  slightly  steeper, $\Gamma= 2.15\pm 0.25$, and even at the CCD  resolution of
the XIS instrument  several lines  from O, Ne, Mg and Si are clearly detected (see
Fig.~\ref{fig:soft.ps}; black   and red data points).  Taking into account the
lower count statistics of the soft X-ray spectrum, we decided to use  a finer
binning for the XIS data adopting  a  minimum  of 50 counts per bin. The fit
statistic of the continuum model is now $\chi^2/dof= 746.4/676$.  We first added 
to the continuum  model several narrow ($\sigma=10$ eV) Gaussian lines, allowing  
also all the continuum parameters   to vary; overall upon including 6 lines the 
fit statistic improves and it is now $\chi^2/dof= 662.8/676$ ($\Delta \chi^2=-83$
for 12 dof). We note that the  $\Gamma$ of the soft power-law is now similar to the
primary power-law component,  we   thus   constrained the two photon indices   to
be the same.\\

In Table\,\ref{table:soft_lines}, we list all the detected lines with their
properties, statistics and possible identification, which point toward emission
from lighter elements  in particular $2\rightarrow1$  transition of H- and He-like
O,  Mg and Si.  Though some of the soft X-ray emission lines are not detected with
high statistical significance (e.g. for {Mg\,\textsc{xi}} we have $\Delta
\chi^2=5.2$),    their detection and interpretation is   in agreement  with the
results obtained with   the deep \chandra\ LETG observation \citep{445chandra}.  
The \chandra\ spectrum confirms the  identification of the feature  detected at
$\sim$0.88 keV with  {O\,\textsc{viii}}  RRC.   This feature, along with the OVII
RRC at $E\sim 0.74$ keV, are resolved by the \chandra\ LETG    and have measured
widths which imply that the emitting gas is photo- rather than
collisionally-ionised. \\ 

Since with the \suzaku\  XIS CCD resolution, we cannot resolve the line triplets
and  we cannot establish with high accuracy the identification of some of the
lines, and taking into account that  \sorg\ is not highly variable, we adopted for
the soft X-ray emission the best fit model obtained with the \chandra\ LETG data. 
This model includes  two grids  of photoionized emission models (with log
$\xi_1\sim 1.82$ erg cm s$^{-1}$ and log $\xi_2\sim3.0$ erg cm s$^{-1}$) generated
by  {\sc xstar} \citep{xstar}, which assumes a $\Gamma\sim 2$  illuminating
continuum and a turbulence velocity of $\sigma_{\rm {v}}=100 $ km/s   and a column
density for  the emitter of N$_\mathrm{H}=10^{22}$\nh. We note that the column
density of the emitter cannot be directly measured from the spectrum because it is
degenerate with its covering factor and thus  the normalisation of this component.
We thus kept the value fixed to the one adopted with the \chandra\ LETG analysis. 
We applied this best fit model to the XIS soft spectrum, keeping the  abundances
fixed to the values measured with the LETG spectrum and   allowing  only the
normalisations and  the photon index   to  vary. This model is now a good
description of the soft spectrum  ($\chi^2/dof= 696.6/674$) and no strong residuals
are present below 3 keV. As a final test, we allowed  the ionisation parameter  of
the two emitters  to vary and found  a good agreement between the \suzaku\ and
\chandra\ best-fit  (log $\xi_1=1.95_{-0.08}^{+0.14}$ erg cm s$^{-1}$ and log
$\xi_2=3.17_{-0.23}^{+0.21}$ erg cm s$^{-1}$). We note  that the  presence of the
higher ionisation emitter is not statistically required ($\Delta\chi^2=3$),  and a
good fit is found with  a single zone with log $ \xi=1.97_{-0.12}^{+0.24}$ erg cm
s$^{-1}$.  We note also that  there is still a line like residual at $\sim 2.3$
keV, which cannot be modelled with these two components. We thus included in the
model an additional Gaussian line, the line is found to be  $E=2.33\pm0.05 $ keV 
and could be associated with S  $K\alpha$. Finally,  we note that both the
ionisation parameters and the  fluxes of the ionised emitters measured with
\suzaku\ are consistent with the one measured with the \chandra\ LETG data. \\

 The extraction region  of the \suzaku\ XIS spectra includes the Narrow Line QSO  
(1WGA J2223.7-0206), which was firstly detected by ROSAT and which is located at
about 1.3' from the \sorg.  Its  X-ray spectrum obtained with \xmm\ was analysed 
and discussed by \citet{NLSY1}. The observed X-ray flux  (F (0.2-10 keV)$\sim
3\times 10^{-13}$\flux, with 80\% emitted below 2 keV)  was found to be comparable
in the soft X-ray band to the emission of  \sorg\ and could thus, in principle,
strongly affect the \suzaku\ spectrum. \\

We note however that a  contemporaneous \swift\ observation (of about 10 ks) of
\sorg\ did not detect 1WGA J2223.7-0206 in the field of view of the \swift-XRT,
suggesting that 1WGA J2223.7-0206 was much fainter during the Suzaku observation. 
We estimated  an upper limit on its soft X-ray flux of $\sim 2\times
10^{-14}$\flux. We also note that the observed soft X-ray fluxes of \sorg\ measured
with \chandra\ and \suzaku\  are   similar:   F ${(0.5-2)\,\mathrm {keV}}\sim 
1.9\times 10^{-13}$\flux and F ${(0.5-2)\,\mathrm {keV}}  \sim 2.3\times
10^{-13}$\flux  for the \chandra\ and \suzaku\ observation respectively. This   
suggests  that the X-ray emission of  1WGA J2223.7-0206 was not comparable to
\sorg, indeed the  the roll angle of the   \chandra\ observation of  \sorg\ was
specifically set  in order to avoid the contamination from  Narrow Line QSO.\\
  
As a last check, we searched  also the \chandra\ source catalog \citep{Evans} 
and the  Chandra   XAssist    source list \citep{xassist} for   X-ray bright
sources within the XIS extraction radius.  Three sources are detected with a 0.3--8
keV  flux greater than $9\times 10^{-15}$\flux  and  1WGA J2223.7-0206 is the
brightest among them  with a 0.5--8 keV  flux of about  $5\times 10^{-14}$\flux. 
Thus we inspected the \chandra\ ACIS-S spectrum of 1 WGA J2223.7-0206, which has 
$\sim 300 $ net counts. In order to derive an  estimate of the soft X-ray flux, we 
fitted  the \chandra\ data  with a  single absorbed power law component 
($\Gamma\sim 1.7$, N$_\mathrm{H}\sim 3\times 10^{21}$ \nh). We  found that also
during the \chandra\ observation   1WGA J2223.7-0206 was fainter than  during the
\xmm\ pointing  and its    0.5--2 keV  observed flux was $1.6\times 10^{-14}$\flux.
We thus conclude that the contamination of this second AGN is minimal. \\ 

\subsection{Modelling the Fe K$\alpha$  line and the high energy emission}
We then considered the hard X-ray emission of \sorg\,  using for the soft X-ray
emission   a single photoionised plasma plus a  Gaussian emission line at $\sim
2.33$ keV as described above   and keeping  the ionisation parameter fixed to the
best fit value. For the remainder of the analysis   we  used  again  the XIS data
grouped with a  minimum  of 100 counts per bin. We examined simultaneously the
\suzaku\ XIS (0.4--10 keV)  and HXD-PIN data (15.-- 65. keV) and the \swift-BAT,
setting the cross-normalization factor  between the HXD and the XIS-FI spectra  to
1.16, and allowing the   cross-normalization with the Swift data   to vary, since
the two observations are not simultaneous.  \\

As shown in  Fig.~\ref{fig:fe_line.ps}   the residuals at the  energy of the Fe K
band clearly reveal the presence of a   strong narrow core  at the expected energy
of the Fe K$\alpha$ line  (6.4 keV), while no clear residuals are present at the
energy of the  Fe K$\beta$ line.  To model the Fe K$\alpha$ line we  first added a 
narrow Gaussian line at the energies of Fe K$\alpha$; the inclusion of the  line 
in the model improves the fit by  $\Delta\chi^2= 101$  for 3 degrees of freedom
($\chi^2/dof= 456.3/405$).  The Fe $ \mathrm{K}\alpha$ core has an equivalent width
of $EW= 100 \pm 14 \,\mathrm{eV}$ with respect to the observed continuum, it is
centered at $E=6.383 \pm 0.012 \,\mathrm{keV}$ and has a measured width of $ \sigma
= 34 \pm 30 \,\mathrm{eV}$. As suggested by the residuals a Fe K$\beta$ is not
statistically required, however we found that the upper limit   on its flux is 
$13.6$\% of the Fe K$\alpha$ line flux, consistent with  the theoretical value. The
amount of reflection ($R=0.9\pm 0.4$) is found to be consistent with the observed
EW of the Fe K$\alpha$ line, for an inclination angle $i= 60^\circ$ and $\Gamma\sim
1.8 $. \\

Finally, we note that  in the XIS-FI data, there are still some line-like residuals
red-wards the Fe K$\alpha$ line.   Upon adding a    second  narrow Gaussian line
the fit  only marginally improves ($\chi^2/dof=445.1/403$  corresponding to $\Delta
\chi^2=11$ for 2 dof), significant at 99.6\% confidence from the F-test.  If this
emission line is real the closest candidate for this  feature could be the Compton
shoulder to the Fe K$\alpha$ line. We note however that its energy ($E=6.05\pm
0.11$ keV) is    slightly  lower than the expected value of the first scattering
peak (E$\sim 6.24$ keV).   An alternative possibility is that this line is a
redwing of a possible  relativistic disk-line. Thus  we    replaced the Gaussian  
with   relativistic diskline component   (DISKLINE  in XSPEC;
\citealp{Fabian1989});    this code models a line  profile from an accretion disk
around a Schwarzschild black hole.  The main parameters of this model are the inner
and outer radii of the emitting region on the disk, and its inclination. The  disk
radial emissivity is assumed to be a  power-law, in the form of $r^{-q}$. For the
fit we  fixed  the emissivity   $q=3$ and the angle to $60^{\circ}$. The fit
statistic is similar to the  Gaussian profile ($\chi^2/dof=446.6/402$), and we
found that the best fit parameters of this diskline corresponds to emission from an
annulus at $\sim 100$ R$_{\rm g}$  (with $R_{\rm g}=GM/c^2$) and with an EW$= 75
\pm 40$ eV. The energy of this putative line, although not well constrained, is 
consistent with the ionised Fe line ($E=6.64\pm 0.16$ keV). Given, the lower
statistical significance and the uncertainties on the energy centroid of this
possible emission line,  we will not discuss it any further.\\

We then  replaced the \,\textsc{pexrav}  and Gaussian components with a more
updated model  for the Compton reflection   off an optically-thick photoionized
slab of gas, which includes the Fe K  emission line (\textsc{reflionx}; 
\citealp{Ross05,Ross99}). We  assumed Solar abundances, and we found the  fit is
equally good   ($\chi^2/dof=453.8/405$). As expected the ionisation of the
reflector is found to be low, $log\xi<1.74$ \logxi, in agreement with the measured
energy centroid of the iron K$\alpha$ emission line being close to the   value for
neutral iron (or less ionised than Fe \,\textsc{xvii}). We thus fixed the
ionisation parameter to  $\xi=10$ \logxi, which is the lower boundary for the
\,\textsc{reflionx} model. We note   that  the residuals at $\sim 6.05$ keV are
still present, we thus keep in the model the additional redshifted emission line.
The parameters of the absorber are  consistent with the values obtained with   the
\,\textsc{pexrav} model (N$_\mathrm{H1} = 1.0^{+0.1}_{-0.2}\times10^{23}$ \nh\   and
N$_\mathrm{H2} = 3.2^{+0.3}_{-0.3}\times 10^{23}$ \nh,  $f_\mathrm{cov} =
0.78^{+0.02}_{-0.01}$). The photon index is now $\Gamma=1.74^{+0.06}_{-0.05}$. 

  \begin{table} \caption{Summary of the   neutral partial covering  absorber  model. $^a$The ionisation of
the reflector has been fixed to the minimum value allowed by the model; if left free to vary the
upper limit is found to be 55 erg cm s$^{-1}$. The ionisation parameters of the soft X-ray plasma is fixed
to  log $\xi=1.97$ erg cm s$^{-1}$. Fluxes are corrected only for Galactic absorption, while the
luminosities are corrected  for rest frame absorption.
\label{tab:bestfit_neut}
}
 \begin{tabular}{lcc}
\hline
 Model Component  &  Parameter  &  Value \\ 
 \hline
&&   \\
Power-law &$\Gamma$&$1.74_{-0.05}^{+0.06}$ \\
& Normalisation & $3.82_{-0.48}^{+0.63}\times 10^{-3}$ \\

Scattered Component  &Normalisation &$9.9_{-0.5}^{+0.5}\times 10^{-5}$\\
Absorber   & N$_\mathrm{H1 }$& $1.0^{+0.1}_{-0.2}\times 10^{23}$ \nh \\
 Absorber  & N$_\mathrm{H2 }$& $3.2_{-0.3}^{+0.3}\times 10^{23}$ \nh\\
          &f$_{\mathrm{cov}}$&$0.78^{+0.02}_{-0.01}$   \\
 Ionised reflection &$ \xi$&$10^a$ \logxi\\
  & Normalisation &$1.20^{+0.12}_{-0.13}\times 10^{-5}$ \\
 Ionised emission  & Normalisation & $2.2^{+0.6}_{-0.6}\times 10^{-6}$
 \\
    &$\chi^2/dof$&454/406\\
&F $_{(0.5-2)\mathrm {keV}}$ &$2.71\times 10^{-13}$\flux\\
&F$_{(2-10)\mathrm {keV}}$  &$7.01\times 10^{-12}$\flux\\
&L $_{(0.5-2)\mathrm {keV}}$  &$7.1\times 10^{43}$\lum\\
&L$_{(2-10)\mathrm {keV}}$  &$1.2 \times 10^{44}$\lum\\
&L$_{(14-150)\mathrm {keV}}$  &$3 \times 10^{44}$\lum\\
 \hline
\end{tabular}
\end{table}

\begin{figure*}
\begin{center}
 \resizebox{0.65\textwidth}{!}{
\rotatebox{-90}{
\includegraphics{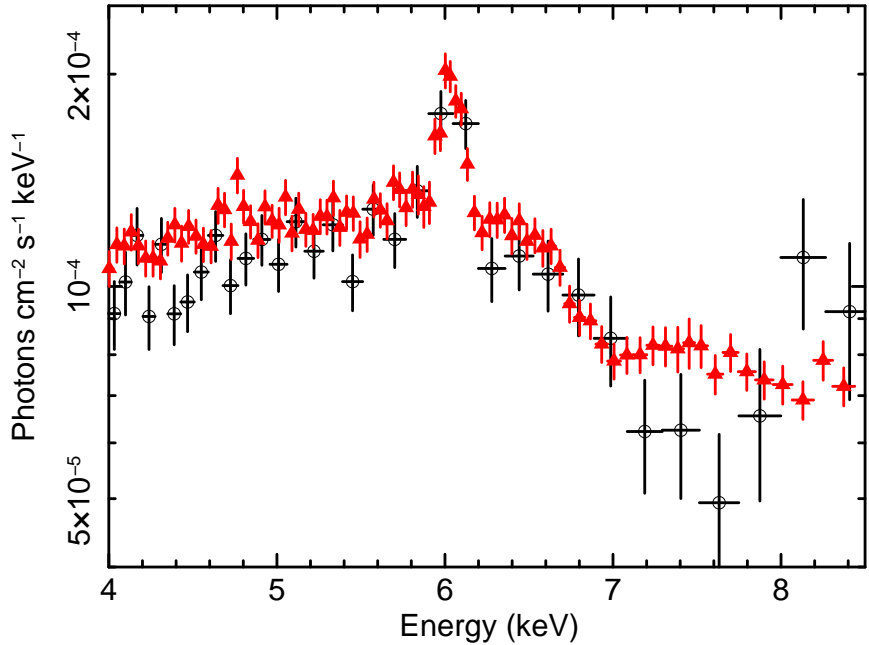}
}}   
\caption {\chandra\  (black circles in the electronic version)  and \suzaku\   
XIS-FI (red triangles in the electronic version) spectra   in the 4--8.5 keV band
folded against a simple power law continuum with the normalisations free to
vary.    In the high energy part, the spectra show a sharper absorption feature in
\chandra\ spectrum compared to a more shallow and possibly broader  drop in the
\suzaku\ data. 
\label{fig:chandra.ps}
}
\end{center}
\end{figure*} 
 
\subsection{The X-ray absorber}
Assuming that the absorber is neutral the broadband X-ray emission of \sorg\
requires the presence of two absorbers one fully covering and one only partial
covering. The best fit parameters of this  model, which  we now consider our
best-fit neutral absorber model, are listed in Table\,\ref{tab:bestfit_neut}. This
dual absorber  plus reflection model is a good  phenomenological description  of
the broad band X-ray emission of \sorg\,; however it may be  too simple an
approximation of  a more complex absorber. We note also that, taking into account
the optical classification of \sorg\ as type 1 AGN,   the X-ray absorber is not
likely to be  a neutral absorber covering a large fraction of the nuclear source as
derived with the above model. A possibility is that the absorber is mildly ionised
and thus it is partially transparent  and  not efficient in absorbing the optical 
and soft X-ray emission.\\

To test this scenario, we then replaced  both the partial  and the fully covering
neutral absorbers  with a  photoionised absorber, the latter is made using a
multiplicative grid of absorption model    generated with the  {\sc xstar} code
(\citealp{xstar}). For simplicity  we modelled the possible relativistic emission
line with an additive Gaussian component  and we allowed to vary the Galactic
absorption (N$_\mathrm{H}  = (0.63_{-0.23}^{+0.22})\times 10^{21}$ \nh). At first
we assumed a zero  outflow velocity of the absorber and we tested a single zone of
absorption. The absorber is found to be mildly ionised  ($log\xi  = 1.10_{-0.24}^
{+0.10}$ \logxi) and with a column density similar to the neutral absorber
N$_\mathrm{H}  = (1.89_{-0.06}^{+0.09})\times 10^{23}$ \nh. We note  also that the 
fit  marginally improves    with respect to the neutral absorber
($\chi^2/dof=443.1/406$) and overall the model is able to reproduce the curvature
of the spectrum.  The best fit parameters of this  model   are listed in
Table\,\ref{tab:bestfit_ion}. \\ 

A similar ionised absorber was also found with  the \chandra\ observation ($log\xi 
\sim  1.4 $ \logxi, N$_\mathrm{H}  \sim  1.85 \times 10^{23}$ \nh); however, the 
LETG spectrum of \sorg\   suggested that this  absorber is outflowing with a
$v_{out}\sim0.034c$, indeed this solution was statistically preferred to a zero
outflow velocity ($\Delta C=22$, \citealt{445chandra}). Thus we allowed the
absorber to be outflowing, but   this   does not statistically improve  the fit
($\Delta \chi^2=2$).  The    parameters of this absorber (N$_\mathrm{H}$ and $log
\xi$) are found to be similar to the case with no net velocity shift and   albeit
it is not well constrained the outflowing velocity is found to be  slightly lower
than the \chandra\  one ($v_\mathrm{out}< 0.01 c$). \\

 \begin{table}
\caption{Summary of the  ionised absorber  model.  $^a$The ionisation of the
reflector has been fixed to the minimum value allowed by the model.  Fluxes are corrected only for Galactic absorption, while the luminosity are
corrected  for rest frame absorption. The ionisation of the
emitter  has been fixed to best fit value.
\label{tab:bestfit_ion}
}
 \begin{tabular}{lcc}
\hline
 Model Component  & Parameter  &  Value \\
 \hline
 &&   \\
 
Power-law &$\Gamma$&$1.85_{-0.04}^{+0.05}$ \\
& Normalisation & $4.79_{-0.69}^{+0.42}\times 10^{-3}$ \\
Scattered Component  &Normalisation &$10.2_{-0.6}^{+0.6}\times 10^{-5}$\\
Absorber   & N$_\mathrm{H1 }$& $1.89^{+0.09}_{-0.06}\times 10^{23}$ \nh \\
          &$log\xi$ & $1.10_{-0.24}^{+0.10}$ \logxi \\
 
Ionised reflection &$ \xi$&$10^a$ \logxi\\
  & Normalisation &$1.26^{+0.15}_{-0.14}\times 10^{-5}$ \\
 
 Ionised emission &  Normalisation & $2.4^{+0.8}_{-0.7}\times 10^{-6}$\\
 
    &$\chi^2/dof$&443/406\\
 &F $_{(0.5-2)\mathrm {keV}}$ &$2.86\times 10^{-13}$\flux\\
&F$_{(2-10)\mathrm {keV}}$  &$7.03\times 10^{-12}$\flux\\
&L $_{(0.5-2)\mathrm {keV}}$  &$8.5\times 10^{43}$\lum\\
&L$_{(2-10)\mathrm {keV}}$  &$1.3 \times 10^{44}$\lum\\
&L$_{(14-150)\mathrm {keV}}$  &$3 \times 10^{44}$\lum\\
 \hline
\end{tabular}
\end{table}

To further investigate  the apparent discrepancy between the \chandra\ and
\suzaku's results we performed a joint fit of the two observations. Though the
source is not highly variable we allow  the relative normalisations of the primary
continuum and the parameters of the ionised absorber to vary. With this test  we
found that the intrinsic emission is slightly brighter during the \suzaku\
observation ($\sim 10\%$). As seen with the independent fit the N$_\mathrm{H}$ and
$log \xi$  of the ionised absorbers are found to be consistent within the two
observations ($\Gamma\sim 1.73^{+0.22}_{-0.19}$, N$_\mathrm{H }\sim
1.85^{+0.09}_{-0.11}\times 10^{23}$ \nh, $\xi\sim 1.4_{-0.12}^{+0.20}$ \logxi  and
$\Gamma\sim 1.85_{-0.04}^{+0.05}$, N$_\mathrm{H }\sim 1.89 ^{+0.09}_{-0.06}\times
10^{23}$ \nh, $\xi\sim 1.1_{-0.24}^{+0.10}$ \logxi from the  \chandra\ and \suzaku\
best fit  respectively; see also Table 3 of \citealt{445chandra}) but  not the
outflowing velocities.  The comparison between the \chandra\  and \suzaku\ data is 
shown in Fig.~\ref{fig:chandra.ps}; in the \suzaku\ data  the drop at high energies
appears to be  broader and less deep.   We note that there is also a   possible
hint of a higher curvature of the \chandra\ spectrum,  indeed the \chandra\
spectrum  is below the \suzaku\ data also  between 4--6 keV. This  could be a
signature of a  variation of the X-ray absorber,  being less transparent during the
\chandra\ observation.   As we show below, with the present data and taking into
account the complexity of the model we cannot confirm or rule out a modest
variability of the absorber. \\

\begin{figure*}
\begin{center}
 \resizebox{0.65\textwidth}{!}{
\rotatebox{-90}{
\includegraphics{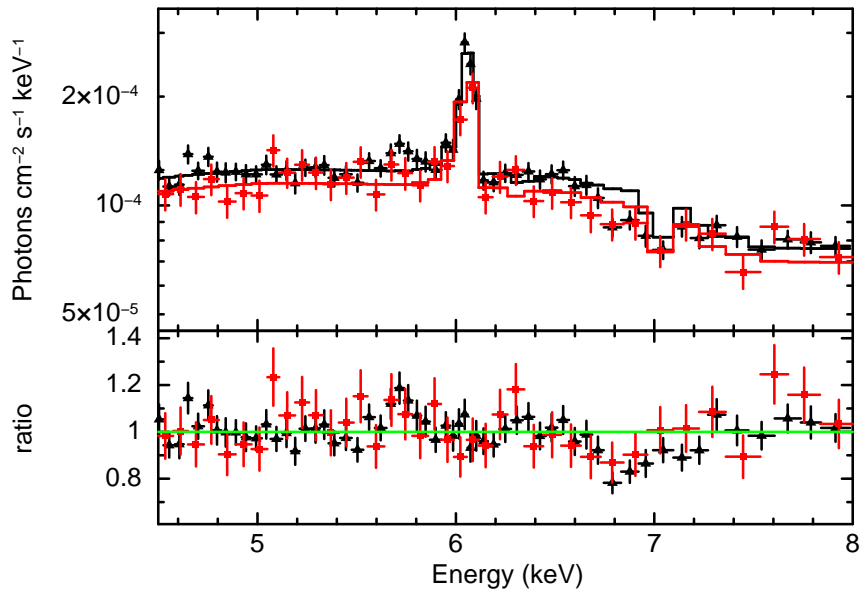}
}}
\caption{Upper panel: \suzaku\    XIS-FI (filled triangles, black in the electronic
version), \xmm\ EPIC-pn (filled squares, red in the electronic version) band folded
against a single mildly ionised absorber    model with the  outflowing velocity
fixed to the \chandra\ best-fit value.   Lower panel: data/model ratio to the above
single ionised absorber. The XIS-FI and EPIC-pn  residuals are consistent  with
each other. They both show a  deficit of counts around $\sim 6.8$ keV (observed
frame, corresponding to $\sim 7.2$ keV in the rest-frame) . 
\label{fig:figura_xmm.ps}
} 
\end{center}
\end{figure*} 

As a final check  we   inspected the previous \xmm\ observation, indeed  also in
that observation there was a hint of a possible absorption feature at ~6.9 keV
\citep{S07}, though the underlying continuum shape was slightly different.  In
particular, lacking the high energy data,  the  amount of reflection could not be
constrained. Taking into account, that the source did not strongly  vary (in shape
and flux), between the two observations, we  then  simultaneously fitted the \xmm\
and \suzaku\  spectra allowing  the cross-normalization to vary and we tested a
single ionised absorber (with no outflowing velocity).   We found that the \xmm\
spectrum is remarkably in  agreement with the \suzaku\ one. In particular we note
that the residuals are  similar also   redward the Fe K$\alpha$ emission  line.
However the short net exposure time of the \xmm\ observation ($\sim 15 $ ksec)
prevents us from further investigating the presence of a possible redwing of a
relativistic Fe K$\alpha$ line.     \\

Two possible scenarios could    explain the  observed differences between the
\suzaku\  and the \chandra\  observations, the first  is that the absorber has
indeed varied, with a lower ionisation and outflowing velocity during the \suzaku\ 
pointing. A second possibility is that the broader drop seen in \suzaku\   is due
to the presence of a more complex and possibly multi-phase absorber. In order to
test the second scenario we fixed  the outflowing velocity of the ionised absorber
to the one seen with the \chandra\  observation. The fit is statistically worse
($\chi^2= 496.7/406$ corresponding to a $\Delta\chi^2=53$)  and clear residuals
are  present, both in the \xmm\  and \suzaku\  spectra,  at $\sim 7.2 $ keV ($\sim
6.8 $ keV in the observed frame), which are reminiscent of a possible absorption
feature. Indeed, the \chandra\  absorption feature appears to be slightly narrower
and more blueshifted, compared to the drop in the \suzaku\  data. Forcing the low
ionisation absorber fitted to the \suzaku\  data to have the same outflow velocity
as inferred from the \chandra\ observation, then results in a deficit of counts
around 7 keV (observed frame) in the \suzaku\ and \xmm\ spectra, when compared to
the  \chandra\  model  (see Fig.~\ref{fig:figura_xmm.ps}). \\

The deficit could then be modelled with an additional absorption line in the Suzaku
data, perhaps arising from a higher ionisation absorber. As a first test, using
only the \suzaku\ data, we included in the model an additional  inverted
Gaussian,   statistically the   fit  is similar  with respect to the single ionised
absorber with no outflow velocity  ($\chi^2/dof=450.9/403$). The energy of this
line is found to be $E=7.30\pm 0.05$ keV  (6.9 keV in the observed frame) and the
$EW =62^{+23}_{-22} $eV,  which would imply a column density  of the absorber  of
about  N$_\mathrm{H }\sim 10^{23}$\nh. The closest candidate  for this absorption
feature is  the  $1\rightarrow 2 $ transition	transition of {Fe\,\textsc{xxvi}}
($E = 6.97$ keV)  blueshifted by $v \sim 0.05\;c $, while if the absorption is
associated with lower ionisation 1s-2p   of  {Fe\,\textsc{xxv}} ($E = 6.7$ keV), 
the corresponding blueshift will be  higher  ($v \sim 0.09\;  c$).   \\

 \begin{figure*}
\begin{center}
 \resizebox{0.65\textwidth}{!}{
\rotatebox{-90}{
\includegraphics{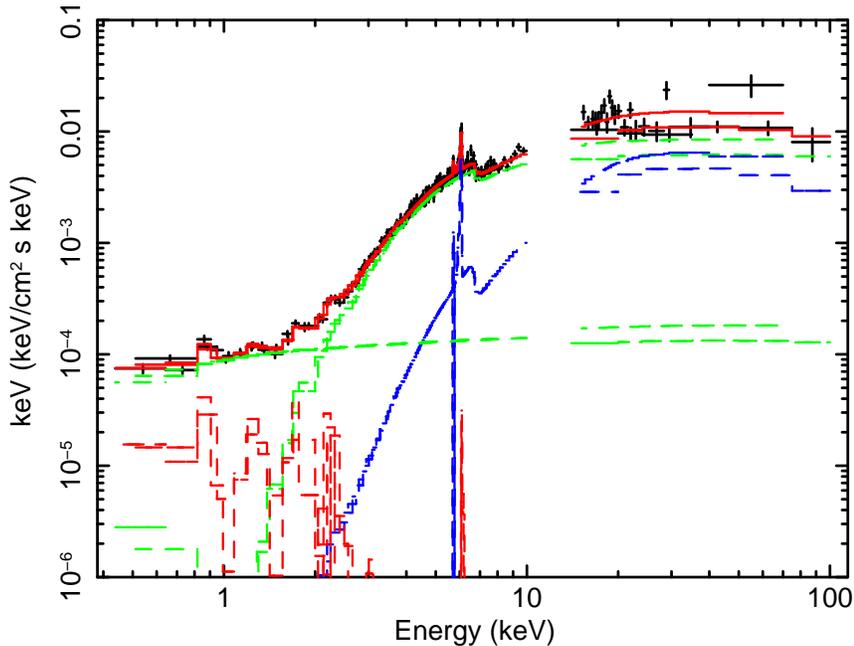}
}}   
 \caption{Suzaku  and Swift-BAT  spectra (black data points) of \sorg. Data have
been rebinned for plotting purposes.   The AGN continuum model is composed of a  
primary power-law component  transmitted trough an ionised absorber  and a
scattered power-law component.   The  narrow Fe K$\alpha$ line  and the reflection
component are modelled with the reflionx model.  The possible relativistic
component is modelled with a single redshifted Gaussian line. The soft X-ray
emission lines are accounted for with an ionised emitter. 
\label{fig:fig7}
}
\end{center}
\end{figure*}

We  then tested   a more complex model for the absorber including a second ionised
absorber and leaving free to vary the outflow velocity of this absorber, while the 
outflow velocity of the mildly ionised  absorber  was fixed to    best fit value
found with \chandra\  ($v=0.034c$). As for the mildly ionised absorber we used a
grid created with {\sc xstar}, for this grid we assumed  again solar abundances,
a   $\Gamma\sim 2$  illuminating continuum and,  taking into account the apparent
broadening of the  absorption feature, we assumed a turbulent velocity of 3000 km
s$^{-1}$.   The addition of this second absorber  does not statistically improve 
the fit with respect to the scenario with  a single ionised  absorber with no net
velocity shift ($\chi^2/dof=453/403$).  This absorber if  found to be fast
outflowing $v\sim 0.04c$, highly ionised log $\xi> 4.7$ \logxi and high column
density N$_\mathrm{H }\sim 10^{24}$  cm$^{-2}$  (N$_\mathrm{H }=2.8^{+0.8}_{-2.6}\times 
10^{24}$ \nh).   It is important to note that  the  significance of this second
absorber is hindered by the choice of the underlying continuum model and more
importantly by the choice of  outflow velocity of the mildly ionised absorber. \\

The simplest interpretation  is  that there is  a modest variability of the mildly
ionised absorber, while we cannot rule out the presence of a second highly ionised
absorber.  To distinguish between these scenarios we need  higher spectral
resolution  observations, such as the one that will be provided with the ASTRO-H
calorimeter.     
 
\section{Discussion and conclusions}
To summarise the \suzaku\ data confirm the complexity of the X-ray emission of
\sorg. The soft X-ray spectrum   is dominated by several emission lines, which
require the presence of at least one ionised emitter with $log \xi \sim 1.97$
\logxi and which is in agreement with   the results obtained with our  deep
\chandra\ observation.  The primary X-ray continuum is strongly   obscured   by an
absorber with a column density of $N_{\mathrm H} =2-3 \times 10^{23}$\nh, which
could be either  a neutral partial covering absorber or a mildly ionised absorber. 
Independently from the model assumed for the absorber, the broadband \suzaku\ 
spectrum allowed us to detect a relatively strong reflection component. \\

Overall, the X-ray spectrum of 3C 445 is remarkably similar to a Seyfert 2, which
could be at odds with its classification as a type 1 AGN.  The main characteristics
resembling a Seyfert 2 are  the presence of soft X-ray emission lines as  well as
the presence of a high column density X-ray absorber.  As we will discuss in the
next sections two competitive scenarios could explain the X-ray emission of \sorg, 
both  requiring that our line of sight is not completely blocked towards the
central engine since we have evidence that we can see the emission from the Broad
Line Regions \citep{Eracleous94}. We will show that our deep \suzaku\ observation
combined with a recent  \chandra\  observation,  with the high-resolution grating
LETG, strongly suggest that both the absorption and the soft X-ray lines originate 
within the putative torus  and we are not seeing the source through  a uniform and 
cold absorber.  

\subsection{The Soft X-ray emission, similarity and differences to Seyfert 2s}

The \suzaku\  spectrum confirms the presence of several soft X-ray emission lines, as previously detected with
the \xmm\ observation, from Oxygen, Neon, Magnesium and Silicon.  In particular, we
detected a line at $\sim 0.88$ keV, which if associated with the   {O
\,\textsc{viii}} RRC would strongly    implies emission from a photoionized plasma 
as seen in Compton-thin Seyfert galaxies \citep{Bianchi06,Guainazzi07}.  As already
shown with the \xmm\ observation  a physical model for the observed  soft X-ray
emission  requires the presence of one or two ionised emitters, with ionisation
levels within the range of  obscured radio-quiet AGN \citep{Guainazzi07}. \\

The limited   spectral resolution  of the XIS data  does not allow us to  resolve
the lines, thus using  only the \suzaku\ data   we cannot measure the density of
this plasma and place strong constraints on the location of the emitter. However, 
as \sorg\ is rather constant in flux, we could use the results obtained with  the
long \chandra\ LETG observation, indeed  assuming the same abundances  we found
that the soft X-ray emission  can be similarly described with two ionised emitters
with the same ionisation (log $\xi_1\sim 1.95$ erg cm s$^{-1}$ and log $\xi_2\sim
3.17$ erg cm s$^{-1}$) and luminosity as during the \chandra\ observation.    The
\chandra\ data provided also the first measurement  of the densities ($n_e>10^{10}$
cm$^{-3}$)   and distance ($R\sim 0.01-0.1$ pc; \citealt{445chandra}) of these soft
X-ray emitters, which are suggestive of a location within    the putative torus and
reminiscent of the Broad Line Region (BLR).  Furthermore, in the \chandra\ data
several lines were  resolved into their   forbidden and intercombination line
components, and  the velocity widths of the O \,\textsc{vii} and O\,\textsc{viii}
emission lines were  determined ($v_{\mathrm{FWHM}}\sim 2600$ km s$^{-1}$).
Assuming Keplerian motion, this line  broadening    implies an origin of the gas on
sub-parsec scales \citep{445chandra}.  \sorg\ is not an isolated example, indeed
there are other well known cases of Seyfert  1s where the soft X-ray emission lines
appear to be produced in the BLR (e.g. MKN 841, \citealt{Longinotti2010}; Mrk 335,
\citealt{Longinotti2008},   NGC4051, \citealt{Ogle2004}; NGC 5548,
\citealt{Steenbrugge2005}).\\  

Thus the emerging scenario is that the soft X-ray emission \sorg\  is    not
produced in a region coincident with the optical narrow line region, as in obscured
radio quiet Seyfert galaxies.  A possible origin of these lines in the BLR is also
in agreement with the multi-wavelength properties of \sorg, indeed  the  optical
classification as a type 1 AGN, suggests that our line of sight toward the BLR  is 
not completely blocked by a high column density absorber.   This photoionized 
emitter  also resembles the ``warm gas'' observed in  more than 50\% radio quiet
Seyferts 1 \citep{Crenshaw2003},  which acts as a ``warm mirror'' and at the same
time intercepts  the line of sight producing absorption features.   As we will
discuss below these photoionized clouds that we are seeing in emission  might be
associated with the  absorber responsible for the curvature of the X-ray continuum.
\\

\subsection{The X-ray absorber/reflector: a distant reflector?}
The \suzaku\ spectrum, and in particular the data above 10 keV, allow us to  confirm  the
presence of a strong reflection component  and for the first time to provide a measurement of its
intensity. The presence of this component was already suggested with the  \sax\ observation
\citep{Grandi2006,Dadina2007}, indeed \sorg\ was detected with the  PDS instrument.   However,
taking into account the  large field of view (FOV)  of this detector and the presence of a nearby
(z=0.09) and bright cluster A2440  (located only 30' away)  it was not possible to derive strong
constraints on the amount of reflection. The presence of a strong reflection component was also
suggested by the intensity of the Fe K$\alpha$ line  detected in the \sax\ and in  the \xmm\ 
observations (EW$\sim 120$ eV \citealt{S07}), which was consistent with being produced  in
reflection off a  medium with high column density \citep{Turner1997,Mytor}.  \\

Thanks to the smaller FOV of the HXD and more importantly to the \swift-BAT
detection we can now confirm the presence of this component. In particular both the
positional accuracy ($\sim 2'$) and offset with respect to \sorg\ ($\sim 0.5 $') 
reported    54-month Palermo-BAT catalog (\citealt{Cusumano})  are compatible with
the emission from the center of \sorg. As shown in  Fig.~\ref{fig:fig7}, the 
averaged \swift-BAT spectrum is  in agreement with the one obtained with \suzaku.
This suggests that there is no strong variability of the intrinsic emission, but
more importantly that  contamination from the nearby cluster is unlikely to be
present. \\

As shown in section 3.3, the amount of reflection measured is $R\sim 0.9$ implying
a   covering factor of the cold reflector of about 2$\pi$ steradians.  Furthermore,
as we showed in the previous section the  primary X-ray emission of \sorg\ is
obscured by a high column density absorber ($N_H\sim 10^{23}$\nh)  with a covering
factor of about 80\%. In the hypothesis that the absorber is neutral   the
predicted extinction in the optical band would then far exceed the observed
reddening of the source (E$_{B-V}\sim 1$  \citealt{Crenshaw88}). In particular a
scenario where our line of sight intercepts a homogeneous  parsec scale torus is
difficult to reconcile  with the observational evidence,  from the soft X-ray and
optical band, that we are seeing the innermost region of this AGN. On the other
hand   the inclination  angle from the jet ($\sim 60^\circ$) combined with the 
current estimates of the average opening angle  of the ``torus''  implies  that  we
might be looking at the nucleus of 3C 445 on the edge of this putative torus. \\

 Though we note that  data with a high-resolution calorimeter     at the Fe
K$\alpha$ line energy    are  necessary   to resolve the line complex and provide
more stringent constraints on the   line width, the  measured Fe K$\alpha$ width 
$FWHM < 7000 $km s$^{-1}$ is consistent with the width of  the soft X-ray lines 
measured with the \chandra\ LETG data \citep{445chandra} and with the  measured 
FWHM of the  H$\alpha$  ($\sim 6400 $ km s$^{-1}$ \citealt{Eracleous94}) and H
$\beta$ ($\sim 3000$ km/s). The Fe K$\alpha$ line could then be in part  produced
either in the outer part of the accretion disk or in the BLR.     We note however
that the relatively high EW of the Fe K$\alpha$ line implies the presence of a
Compton thick reflector, which is also confirmed  by the detection of a strong
reflection component. Indeed in order  to be produced in transmission the observed
EW of Fe K$\alpha$   requires a higher column density absorber than the  one
measured with lower energy cutoff.\\

   We thus tested the new model for the toroidal
reprocessor\footnote{http://www.mytorus.com/}  \citep{Mytor}, keeping the soft
X-ray emitter modelled with a single ionised zone and we found that the intensity
of the Fe K$\alpha$ line and of the higher energy emission require the presence of
a reprocessor with a column density $N_{\mathrm{H}}\sim 6.5\times 10^{23}$\nh\  viewed at $\sim 60 ^\circ$, which is remarkably in agreement with the inclination
of the system as derived from the radio observations \citep{Eracleous98}. We note
however that the photon index is now harder ($\Gamma\sim 1.5$) and there are some
residuals in the  2--5 keV band  suggesting the  presence of a second absorber. We
thus included a second and  ionised absorber with no net outflowing velocity, 
qualitatively we note that we now have an excellent representation of the overall
curvature of the X-ray emission. While the parameters of the ionised absorber are
similar to the  one derived with the  \,\textsc{reflionx}  model,    we found that
the reprocessor responsible for the Fe K$\alpha$ line has a column density of 
$N_{\mathrm{H}}\sim 1.4\times 10^{24}$\nh\ and as before it is viewed at $\sim 60
^\circ$, the photon index is now $\Gamma\sim 1.8$.  Qualitatively both these tests
show us that a  high column density  mirror responsible for the Fe K$\alpha$ line
and the Compton reflection component is present and it could be  associated with
the part of the putative torus lying close to the plane of the accretion disk. \\

One possibility is that the photoionised emission line clouds,  although located
closer to the central SMBH,  are lifted above the system's equatorial plane;  our
line of sight would then intercept  the   high column density absorber/reflector on
a sub parsec scale but also have  an unobscured view of BLR emission, where   the 
optical and soft X-ray emission lines are produced. The  alternative scenario which
is suggested by  the clumpy and neutral absorber,  required to model the X-ray
emission of \sorg, is that the absorber/reflector is not a uniform ``donut-like''
structure  and as proposed in recent models  \citep{Risaliti2002,Elitzur2008,
Maiolino2010}  is   clumpy and composed by  many small and dense clouds, which
could  extend   further in with respect to putative torus and are not simply
obscuring the BLR but are part of the BLR themselves
\citep{Risa2009a,Risaliti2009b}.    In this hypothesis    our line of sight could
then   intercept    a rather large number of clouds, which absorb/reflect the
primary continuum but at the same time  may also produce the    the broad emission
lines from innermost ionised   clouds. \\

\subsection{The nature of the X-ray absorber: a variable ionised absorber?}
 
As shown in  section 3.4,  an  alternative scenario to a clumpy and neutral
absorber is an ionised absorber. This scenario naturally accounts for the
discrepancy between the optical and the X-ray band, indeed this mildly ionised
absorber  could be similar to the warm absorbers observed in the X-ray spectra of
radio quiet  AGN  \citep{Crenshaw2003,Blustin2005,McKernan2007,Turner2009}, which
appear to be  outflowing with velocity of 100-1000 km/s and could be associated
with the presence of disk winds \citep{King03}. \\

As discussed in a companion paper (\citealt{445chandra}) the presence of an
ionised absorber, with the same ionisation level and column density,  associated
with a disk wind is strongly supported  by the deep observation with  high spectral
resolution which provides more stringent constraints on the velocity of this
absorber  ($v_\mathrm{out}=0.034\pm 0.002 c$) and on the launch radius ($R\sim
10^{16}-10^{17}$cm).  As also shown in that paper    at this distance the likely 
density  of the absorbing gas would be $\sim 10^{10} $ cm$^{-3}$,  which would
imply a $\Delta R\sim 10^{13} $ cm and a $\Delta R/R \sim 10^{-3}$ and thus
suggesting a    highly clumped absorber. It was also suggested that the clumpiness
of this absorber would produce short-timescale variations of the observed column
density, as seen in other Seyferts.  \\

Indeed, the observed difference between the  \chandra\ and \suzaku\ spectra could
be explained with variability of this low ionisation absorber, not in terms of the
covering factor or $N_{\mathrm{H}}$ of the absorber but in terms of its velocity.  
This is not surprising indeed not only   red- and blueshifted absorption lines are
predicted in several theoretical models of failed disk winds
\citep{Proga2004,Sim2010} or of aborted jet  \citep{Ghisellini2004} but also   
these models   predict  the outflows to be  unstable  and  to show variability.  In
particular,    outflows and jets could be produced  intermittently and/or they
could not have enough power to escape the system and  eventually fall back into the
accretion disk. This will  affect the expected signatures  that this  warm gas 
imprints on the primary X-ray emission, which will  produce transient absorption
features and variability of the derived outflowing  velocities  and  their EW as
observed in several sources  (\citealt{Braito07,Porquet2007,Dado05,
Risaliti05,Turner2008,Turner2010,Miller2010}; Lobban et al. 2010).\\

An alternative scenario ascribes the difference between \chandra\ and \suzaku\
spectra  to   a further ionised and fast outflowing  absorption component, detected
only  with the \suzaku\ observation and characterised by a  high ionisation  
and    column density  ($N_{\mathrm {H}} \sim 10^{24}$\nh). In this scenario the
properties of the low ionisation absorber are thus the same as the one derived with
the \chandra\ observation  (i.e., distance $R\sim 10^{16}-10^{17}$cm   and
clumpiness $\Delta R/R \sim 10^{-3}$).  For the high ionisation absorber,  
although in the \suzaku\ data the parameters of this absorber not well
constrained,   we can  derive a  order of magnitude  estimate on its likely
location   from the values measured for the ionisation parameter ($\xi\sim 5$
\logxi), the outflow velocity ($v_{\mathrm {out}}\sim 0.04c$) and the column
density. We can thus   use the relation between these quantities and the 
illuminating continuum luminosity: $L_{\mathrm{ion}}/\xi=nR^2$, where
$L_{\mathrm{ion}}$ is the intrinsic 1--1000 Rydberg   luminosity ($3\times 10^{44}
$ erg s$^{-1}$), assuming  the thickness of the clouds $\Delta R=N_H/n$ is less
than the distance R, $\Delta R/R << 1 $   we  found $R< 5\times 10 ^{15}$cm (or
$\sim$0.001 pc), which points towards an association of this absorbers with a  wind
launched off the disk at a sub-parsec distance from the central BH.\\

We note that both these ionised absorber  scenarios    imply  that we are seeing a
clumpy  and possibly  variable absorber located close to the central X-ray source.
This could be associated with the presence of a disk wind which either is launched
sporadically or it is   highly clumped. However, in the absence of any stringent
constraints on the launch radius we cannot speculate more  as to whether   these
two absorbers are part of a single clumpy wind, where  the lower ionisation
component is associated to higher density clouds confined in the homogeneous highly
ionised flow, or if  the two components  are part of single stratified medium. In
order to    determine whether it is  a single and variable ionised absorber or a
multi-phase wind   would require higher resolution   observations with instruments
such as the calorimeter which will fly with  Astro-H. These observations will
allow   to    establish  the complex nature  and kinematics of this absorber
confirming  the presence of blueshifted absorption lines  from highly ionised iron.

\section{Conclusion}
We have presented the results of a deep \suzaku\ observation of the BLRG \sorg\,
which shows a complex absorbed X-ray spectrum. We confirm the results obtained with
the previous \xmm\ observation which unveiled  the presence of several soft X-ray
emission lines.  The \suzaku\  and \swift\ spectra allowed us to measure a strong
reflection component, which we associate with the presence of a high column density
matter which  is not in the line of sight.  The primary X-ray continuum is strongly
absorbed either by a  partially covering neutral  or a  mildly ionised absorber,
which could be associated with an accretion disk wind. \\

Regarding  the  overall geometry  of \sorg, we know from the radio observations
that we are  seeing the central regions of this AGN at a relatively large
inclination. A plausible scenario is that we are viewing along the edge of the
putative torus through  either a  partially covering neutral absorber or   mildly
ionised absorber, which could be associated with an   equatorial   disk-wind. In
both the scenarios with an ionised or neutral absorber, the  matter needs to be
clumped, such as that the   observer has a direct view of the clouds responsible of
the soft X-ray  and optical lines, which could be in part uplifted with respect to
the  equatorial plane.   A possible schematic diagram for the  geometry of the
inner regions of \sorg\ is presented in Figure 8 of \citealt{445chandra},   a new
addition to that schematic view is that \suzaku\ provided evidence for the presence
also of  a Compton-thick reflector.  We find no evidence that our line of sight
intercepts this Compton-thick absorber which  is responsible for the reflected
component and Fe K$\alpha$ line. This absorber could be either  associated with
denser clouds probably located in the equatorial plane of the  torus or of the
clumpy absorber or the outer part of the disk-wind.   \\

\section*{Acknowledgments}
This research has made use of data obtained from the Suzaku
satellite and data obtained from the High Energy Astrophysics Science
Archive Research Center (HEASARC), provided by NASA's  Goddard Space Flight Center.
VB acknowledge  support from the  UK STFC research council.

\label{lastpage}

\end{document}